# Re-thinking Enrolment in Identity Card Schemes

Dr. Ali M. Al-Khouri

Emirates Identity Authority, United Arab Emirates

**Abstract**
Many countries around the world have initiated national ID card programs in the last decade. These programs are considered of strategic value to governments due to its contribution in enhancing existing identity management systems. Considering the total cost of such programs which goes up to billions of dollars, the success in attaining their objectives is a crucial element in the agendas of political systems in countries worldwide. Our experience in the field shows that many of such projects have been challenged to deliver their primary objectives of population enrolment, and therefore resulted in failing to meet deadlines and keeping up with budgetary constraints.

The purpose of this paper is to explain the finding of a case study action research aimed to introduce a new approach to how population are enrolled in national ID programs. This is achieved through presenting a case study of a business process reengineering initiative undertaken in the UAE national ID program. The scope of this research is limited to the enrolment process within the program. This article also intends to explore the possibilities of significant results with the new proposed enrolment approach with the application of BPR. An overview of the ROI study has been developed to illustrate such efficiencies.

## 1. Introduction

In today's dynamic global business environments, organisations both in public and private sectors are finding themselves under extreme pressure to be more flexible and adaptive to such change. Over the past two decades, organizations adopted business process reengineering (BPR) to respond to such business agility requirements.

This is based on the belief that process is what drives the creation and delivery of an organization's products and services (Evans, 2008). The literature demonstrates that BPR can yield profound and dramatic effects on lowering costs, quality of service delivery and customer satisfaction (Hammer and Champy, 2003; Jeston and Nelis, 2008; Madison, 2005). Thus, it considers it as an important approach to transform operations, and to achieve higher levels of efficiency and effectiveness. In short, BPR is more of a holistic approach to change with a comprehensive attention to process transformation in light of social issues, business strategy, people performance, and enabling technologies.

Many governments around the world have initiated national ID card programs with allocated budgets exceeding multi-billions of dollars. Many of such programs worldwide have been challenged to achieve their core objective of population enrolment (Al-Khouri, 2010). Taking into consideration the strategic objectives of such programs and high budgets, it is deemed necessary that learnings from various implementations are shared between practitioners in the field to address common challenges and learn from best practices. It is the purpose of this paper to contribute to the current body of knowledge and present an action based case study research in one of the most progressive countries in the Middle East. It attempts to present a case study of a process re-engineering project that was implemented in the United Arab Emirates (UAE) national ID program. It also sheds light on the staggering results gained from such an exercise.

This paper is structured as follows. First, a short literature review of the BPR concept is provided. Some background information about the project and the triggering needs for process improvement are discussed next. Some reflections and management consideration areas are discussed afterwards, and it ends with some concluding remarks and possible future research areas related to this topic.

## 2. Literature Review: Business Process Reengineering and NPM

Process reengineering has long history and application as it evolved overtime in various forms to represent a range of activities concerned with the improvement of processes. The reengineering concept goes back in its origins to management theories developed as early as the late eighteenth century, when Frederick Taylor in 1880's proposed process re-engineering to optimize productivity and improve performance. 30 years later, Henri Fayol, instigated the reengineering concept seeking to derive optimum results from available technology resources in a manufacturing environment (Lloyd, 1994).

Some revolutionary thinking was added to the field in the past two decades. For instance, Davenport and Short (1990) presented process re-engineering as the analysis and design of work flows and processes within and between organizations. Extending the work of Porter (1980, 1985, 1990) on competitive advantage, Hammer and Champy (1993) promoted the concept of business process reengineering as a fundamental rethinking and radical redesign of business processes to achieve dramatic improvements in key performance measures e.g., cost, quality, service, and speed (see also Lowenthal, 1994; Talwar, 1993).

The reengineering concept has evolved in the recent years to reconcile with more incremental process management methods such as Total Quality Management; often referred to as TQM (Davenport and Beers, 1995; see also Caron et. Al, 1994; Earl and Khan, 1994). Other researchers have integrated reengineering with other modern management concepts such as knowledge management, empowerment, organization theory, organization control, strategy, and MIS (Earl et al., 1995; Kettinger & Grover, 1995).

Reengineering, in general, questions all assumptions about the way organisations do business and focuses on the *how* and *why* of a business process to introduce major changes to how work is accomplished. In fact, it moves far beyond mere cost cutting or automating a process to make marginal improvements (Cash et al., 1994). According to Davidson (1993), successful reengineering efforts ultimately lead to business transformation. New products, services and customer services appear in the form of improved information flows (ibid).

Several studies pointed out that while the potential payback of reengineering is high, so is its risk of failure and level of disruption to the organizational environment. Introducing radical changes to business processes in an attempt to dramatically improving efficiency and effectiveness is not an easy chore. While many organisations have reported impressive augmentation and accomplishments, many others have failed to achieve their objectives (Davenport, 1993; Keen, 1991). Reengineering in whatever form or name it appears in, seeks to improve the strategic capabilities of an organisation and add value to its stakeholders in some idiosyncratic ways. Strategic capabilities are the means and processes through which value is added, as distinct from the products and services perspectives and their competitive positioning in the marketplace.

In government context, process reengineering has been associated with Public Administration Reform often referred to as New Public Management (NPM), a term used to transform and modernize the public sector. NPM seeks to enhance efficiency of the public sector and the control framework with the hypothesis that more market orientation in the public sector will lead to greater cost-efficiency for governments, without having negative side effects on other objectives and considerations. Dunleavy et al., (2006) defines NPM as a combination of splitting large bureaucracies into smaller, more fragmented ones, competition between different public agencies, and between public agencies and private firms and incentivization on more economic lines (Dunleavy et al., 2006).

It is such concepts that are pushing public sector organizations nowadays to act similar to those in the private sector. Therefore, governments around the world are transforming their mindsets of how they view their citizens and treat them as *customers*, with much emphasis on leveraging technology to building long-term relationships with their citizens. This has raised expectations of *customers'* relentless demands in quality and service in this sector.

The new power and freedom of the *customer* has destroyed many of the organisational assumptions of the early role of government, and placed them as a new powerful stakeholder. So process reengineering in this context is concerned more with facilitating the match between customer needs and organisational capabilities in light of the government roles and responsibilities. Many governments have initiated process re-engineering projects to develop citizen-focused, service oriented government architectures, around the need of the citizens, not those of the government agencies.

By and large, governments nowadays are put under tremendous pressure to strive for operational and financial efficiencies, while building an environment that encourages innovation within the government, in light of population growth, demographic changes, technological and knowledge 'explosions', and increased citizen expectations (Gordon and Milakovich, 2009). The following section outlines the research methodology and it contribution the body of knowledge.

## 3. Research Methodology

The research methodology adopted in this study was a mixed approach of action and case study research. The phenomenon measured in this study was considered to be too complex, and needed to be constructed and measured experimentally, and particular attention was paid to the organisational (and local) idiosyncrasies that permeate all true natural settings.

Action research is defined as "a type of research that focuses on finding a solution to a local problem in a local setting" (Leedy & Ormrod, 2005, p. 114). Action research is a form of applied research where the researcher attempts to develop results or a solution that is of practical value to the people with whom the research is working, and at the same time developing theoretical knowledge. Through direct intervention in problems, the researcher aims to create practical, often emancipatory, outcomes while also aiming to reinform existing theory in the domain studied.

Case study research, on the other hand, is a common qualitative method (Orlikowski & Baroudi, 1991; Alavi & Carlson, 1992). Although there are numerous definitions, Yin (2002) defines the scope of a case study as an empirical inquiry that (1) investigates a contemporary phenomenon within its real-life context, especially when, (2) the boundaries between phenomenon and context are not clearly evident (Yin, 2002).

Action research was found particularly appropriate to investigate and describe the situation, the issues at hand and its context to effect positive change in the situation. Clearly, the case study research method is particularly well-suited to this research, since the objective of our study is the systems in the organization, and our "interest has shifted to organizational rather than technical issues" (Benbasat et al. 1987).

The study development was primarily facilitated by the senior role of the researcher in the examined organization. The study was based on both primary and secondary data. Data were gathered from business documents, technical specifications, annual reports, observation, and both formal and informal discussions with key stakeholders in the organization.

### 3.1 Study Contributions: National Identity Management Systems

Many governments around the world have initiated national Identity management systems. The nature and operating model of these systems make it extremely vulnerable to considerable challenges. The customer base for such programs is basically all resident population in a country setting, and such programs require the physical presence of people to complete the registration process i.e., require the capturing of biometrics.

There are more than 130 countries that have already implemented such systems and many other countries are seriously considering the implementation of such programs. The huge amount of program cost and the complexity of its infrastructure technologies further contribute to it being challenged.

Our previous studies in the field (Al-Khouri, 2010; Al-Khouri, 2007) presented to us that these programs are challenged to meet their primary objective of *population enrolment*. This would in turn have a serious impact on the original implementation time frames and the allocated budgets set by governments. It is also noted that existing literature include very little data about practices about this important and critical field. In fact what makes this study of high contribution is related to the fact that we are not aware of any previous research in this area, which points out the significance of this study. It is our attempt therefore to contribute to the existing body of knowledge and share experiences of such implementations and associated critical insights. This should serve as guidelines for framing their practices in the implementation of similar projects world over.

### 3.2 Research Limitation

It is comprehended that case studies and action research are usually restricted to a single organisation making it difficult to generalise findings, while different researchers may interpret events differently. The research in this study restricted to a single organisation, thus a major limitation of this research is the sample size that limits generalizability.

Having said that, the next section provides an overview of the case study organisation and high level results achieved through business process reengineering of core functions.

## 4. Emirates Identity Authority

The Emirates Identity Authority (Emirates ID) is a federal government authority established in 2004 to develop and manage the implementation of a national identity management infrastructure in the United Arab Emirates.

The organization began to rollout the program in mid 2005 and managed to enrol a population of 1.1 million over a 4 years period. The registration process was widely criticized by the population for being "hectic" and gained the organization with some negative reputation in the media. Long waiting times, complex registration procedures, high turnover rates, increasing costs; all indicated performance setbacks. The organization was in a real dilemma and a change from status was needed.

In late 2009, the Vice Chairman of the Board of Directors at Emirates ID articulated the need for a radical change program to examine lagging performance results. The new appointed Head of Executive Board Committee and the Director General established ambitious goals for drastic improvements throughout the organization, and they paid attention in particular to selected business areas related to the intake and the production cycle capacities at registration centres; where criticism was most. Specific goals included reducing applicant registration and waiting times by at least 50 per cent. The new management team decided to take a system-wide view of the organization and perceive its business as a factory with production lines environment that harnesses the potential of teams.

The workforce at production line processes were evaluated and rewarded based on their performance. In a little over one year, the organization achieved monstrous results; increased intake capacity by 300%, reduced registration time by 80%, reduced applicants waiting time by 1000%, reduced staff turnover by 60%, lifted customer satisfaction by over 52%, increased revenues by 400%, and cutting 300% of overheads. Due to the substantial size and details of the work conducted in the organisation, the discussion in this article was limited to the process re-engineering project performed part of the overall change program, as the next section details.

### 5. The Process Re-Engineering Project

As explained earlier that biggest motive for business process reengineering stemming from key challenges faced by the organization in achieving its objective of enrolling all citizens and residents into the UAE Federal government's "Population Register" program. Key drivers for these challenges were limited daily intake capacity, complex enrollment processes and lack of robust mechanisms to ensure a regular flow of enrollment applications. Hence and once leadership decided that a radical change was needed, it was clear that an enrollment process re-engineering was required. Prior to kicking off the re-engineering project, the team leading the project reviewed leaderships guiding principles for deploying the change process. The four guiding principles mandated by leadership were:

(1) Increased efficiency,
(2) Cost optimization,
(3) Incremental capacity, and
(4) Enhanced customer experience.

These guiding principles thus become the pillars on which the future population enrollment strategy needed to be built on. So, in order to deliver upon each of these guiding principles, the management team studied various options and revamped the end-to-end enrollment process with a specific focus on elimination of bottlenecks and redundant processes. The following subsections will outline key changes implemented addressing each of the guiding principles.

### 5.1 Increased efficiency

The registration process was the obvious reengineering opportunity. Much time was devoted to assessing risks and benefits of various design alternatives. The most important consideration was that the new system needed to be customer centric and driven by customer needs of faster and more convenient registration process. Common complaints from people previously were that of going through long, cumbersome and highly time consuming procedures at registration centres. Accordingly, the new redesigned process yielded the following outcomes:

– Reduction of enrollment processes from a 6 step to a 4 step process, and
– Standardization of biometrics capture technology from maximum of 3 unit workstations to a standard 1 unit workstation (See also Fig. 1).

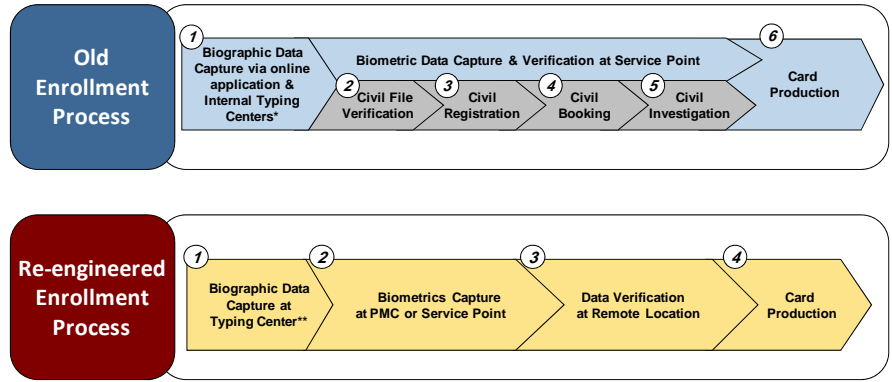

Fig. 1. Old registration process vs new process.

As a result of the above change, the average theoretical time for enrolling each applicant was reduced by 23 minutes per application. Key drivers for this change were a 10 minute reduction in average time to fill-out an ID card application and a 13 minute reduction in average time for biometrics capture and data verification. A key consideration in making the time comparison between old versus new process was that in the past applicants completed their ID card applications online and in the new design, they would pay to have an application completed for them by an authorized typing center.

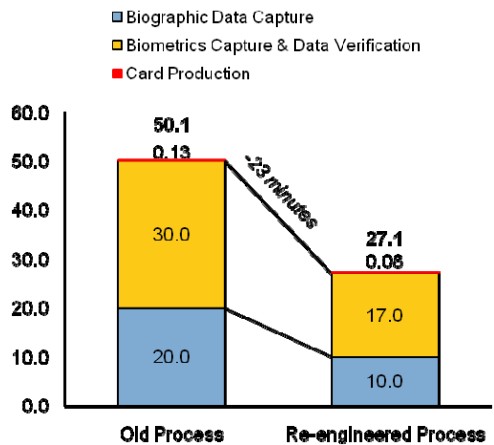

Fig. 2. Time savings in new process.

### 5.2 Cost Optimization

- Shorter processing time per application leading to higher utilization rate and hence increased productivity of enrollment workstation operators.

As a result of the above change, average theoretical overhead (labour) for biometrics capture and data verification was reduced by 30 AED per application. Key drivers for this change was the hypothesis that existing staff would be utilized for biometrics capture and data verification processes. The implications of the above hypothesis was that the average labour cost associated with each workstation operators would remain unchanged at an average cost of 22,000 per month; however each employee would be able to process a greater numbers of applications per day given the 23 minute reduction in lead time. In addition typing centers and outsourcing costs were excluded in order to produce an assessment purely focused on process re-engineering.

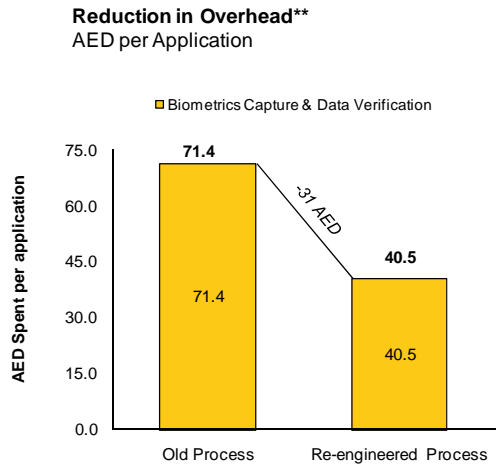

Fig. 3. Cost savings with reengineering.

### 5.3 Incremental Capacity

- Achievement of incremental capacity in biographic capture, biometrics capture and card production processes as of Q3 2010, and
- Additional increases in biographic capture, biometrics capture. Population register processing capabilities and card production processes planned by Q3 2011.

As a result of deploying increased capacity across sub steps of the enrollment process, Emirates ID had the potential of raising end to end daily enrollment throughput to approximately 22,000 applications per day by Q3, 2011.

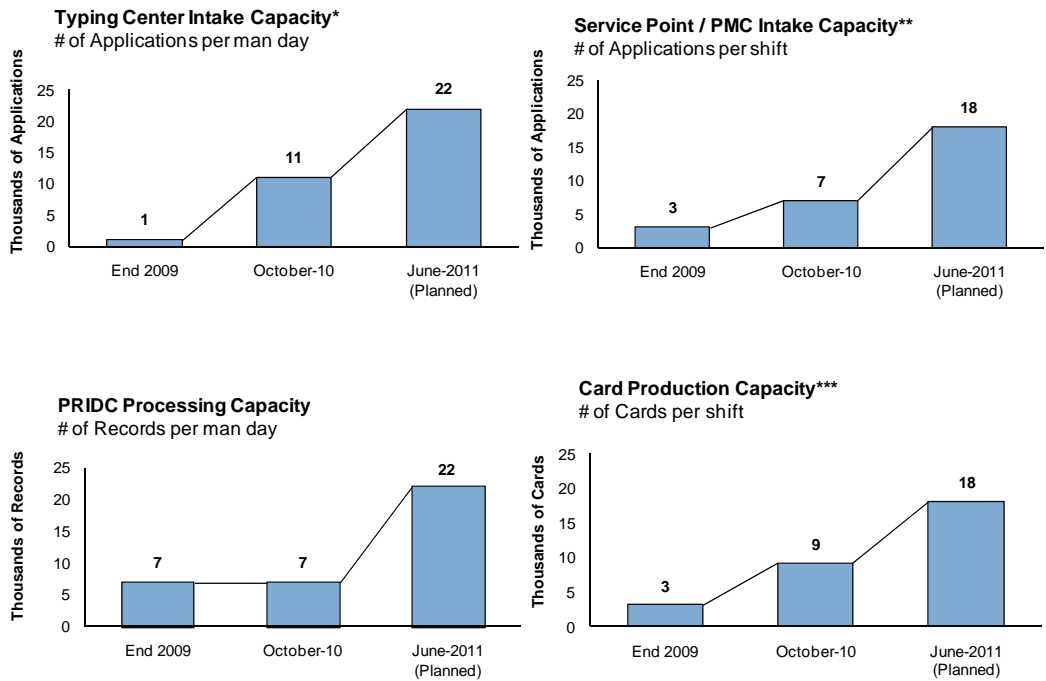

Fig. 4. Capacity development.

### 5.4 Enhanced Customer Experience

In addition to the tangible elements quantified in earlier sections, there were also a number of intangible enhancements that greatly benefit both Emirates ID and its customers. Some of these are mentioned in the diagram below:

| Intangible Elements of Enrollment Processes | Old Process | Re-engineered Process |
|---|:---:|:---:|
| • Presence of applicant for biometrics capture only<br>– Submission of biographic information may be completed by applicant representative<br>– Data verification done at off-site location | ✘ | ✓ |
| • Automation of processes leading to on-line "paper trial", enhanced security, simplified data retrieval and greater "Business Intelligence" | ✘ | ✓ |
| • Ability to manage flow of applicants to enrollment locations (Service Points / PMC's)<br>– Scheduling of appointments for biometrics capture | ✘ | ✓ |
| • "Unified form" (planned) for multiple federal government applications<br>– Simplifying biographic capture process<br>– Benefiting applicant and government entities | ✘ | ✓ |

Fig. 5. Customer service enhancement areas.

### 5. Key Results from Process Reengineering

Once the entire process re-engineering program was completed, Emirates ID was expected to deliver on each of the four guiding principles outlined by leadership listed earlier.

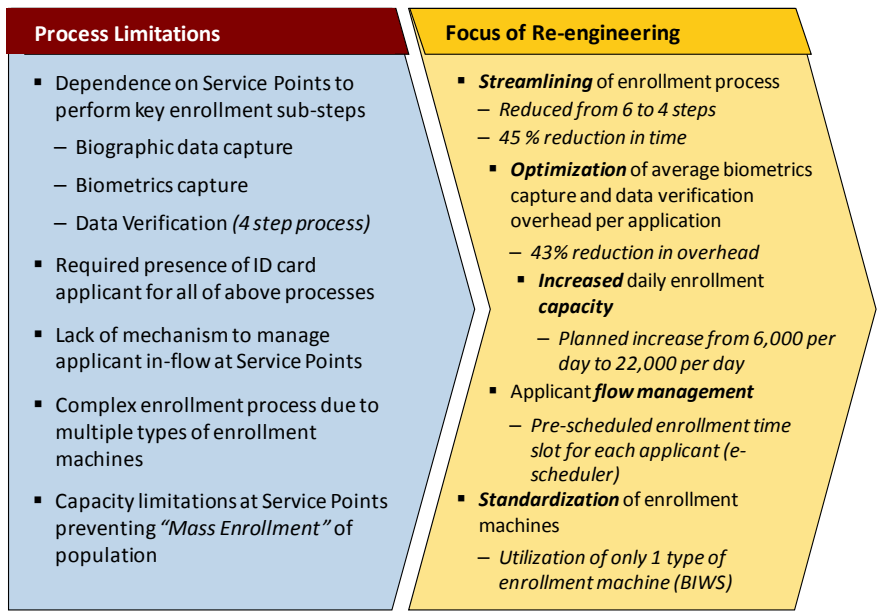

**Process Limitations**
- Dependence on Service Points to perform key enrollment sub-steps
  - Biographic data capture
  - Biometrics capture
  - Data Verification *(4 step process)*
- Required presence of ID card applicant for all of above processes
- Lack of mechanism to manage applicant in-flow at Service Points
- Complex enrollment process due to multiple types of enrollment machines
- Capacity limitations at Service Points preventing *"Mass Enrollment"* of population

**Focus of Re-engineering**
- ***Streamlining*** of enrollment process
  - *Reduced from 6 to 4 steps*
  - *45 % reduction in time*
- ***Optimization*** of average biometrics capture and data verification overhead per application
  - *43% reduction in overhead*
- ***Increased*** daily enrollment ***capacity***
  - *Planned increase from 6,000 per day to 22,000 per day*
- Applicant ***flow management***
  - *Pre-scheduled enrollment time slot for each applicant (e-scheduler)*
- ***Standardization*** of enrollment machines
  - *Utilization of only 1 type of enrollment machine (BIWS)*

\* Does not account for cost of 3rd party contracts which were not part of original re-engineering project
\*\* based on deployment of ~400 BIWS machines currently in inventory (including converted machines)

Fig. 6. Key results from reengineering.

Having done so, not only Emirates ID was prepared to deploy its Mass Enrollment strategy, it was also able to reap the benefits of cost and time savings per applicant enrolled. A high level study of cost and time savings based on potential future scenarios led to the following results.

**5.1 Scenario 1: Linkage of Residence Visa Applications**

The first scenario extrapolated cost and time savings based on linkage of all residence visa applications to the Emirates ID card. If such a linkage were activated across the UAE at the beginning of Q3 2011, Emirates ID was expected to have a constant flow of 15,000 applications per day. Extrapolating this enrolment forecast through the end of 2012 reveals that at that point in time, Emirates ID would have enrolled over 9 million people in its Population Register, and in doing so would save approximately 227 million AED in labour cost and over 117,000 man hours.

| | Cumulative Impact of Overhead and Time Savings (Linkage of Residence Visa Applications) | | | | | | | | |
|---|---|---|---|---|---|---|---|---|---|
| | 2010 | | 2011 | | | | 2012 | | | |
| | Q3 | Q4 | Q1 | Q2 | Q3 | Q4 | Q1 | Q2 | Q3 | Q4 |
| Average New Enrollments / Man Day (Thousands of People) | | 7 | 7 | 7 | 15** | 15** | 15** | 15** | 15** | 15** |
| Additional Enrollment (Millions of People) | | 0.5 | 0.5 | 0.5 | 1.0 | 1.0 | 1.0 | 1.0 | 1.0 | 1.0 |
| Cumulative Enrollment (Millions of People) | 2.3 | 2.8 | 3.3 | 3.7 | 4.7 | 5.7 | 6.7 | 7.7 | 8.7 | 9.7 |
| Est. Total Overhead Savings (Millions of AED) | | 14 | 29 * | 43 | 74 | 104 | 135 | 165 | 196 | 227 |
| Est. Total Time Saving (Thousands of Days Productivity) | | 7.4 | 14.8 | 22.2 | 38.0 | 53.9 | 69.7 | 85.5 | 101.4 | 117.2 |

NOTE : Based on labor cost saving of 31 AED per application and wait time reduction of 23 minutes per application

\* Breakeven on cost of 200 BIWS machines acquired in 2010 to enable re-engineered process (by Q2 2011)
\** All visa applicants required to enroll for Emirates ID card

Fig. 7. ROI scenario 1.

**5.2 Scenario 2: Linkage of Residence Visa Applications & Key Govt. Services**

The second scenario extrapolated cost and time savings based on linkage of all residence visa applications plus key government services (e.g. driver's license, vehicle registrations, etc.) to the Emirates ID card. If such a linkage were activated across the UAE at the beginning of Q3 2011, Emirates ID was expected to have a constant flow of 18,000 applications / day. Extrapolating this enrolment forecast through the end of 2012 reveals that at that point in time Emirates ID would have enrolled over 10 million people in its Population Register, and in doing so would save approximately 265 million AED in labour cost and over 136,000 man hours.

| | Cumulative Impact of Overhead and Time Savings (Linkage of Residence Visa and Govt. Services Applications) | | | | | | | | |
|---|---|---|---|---|---|---|---|---|---|
| | 2010 | | 2011 | | | | 2012 | | | |
| | Q3 | Q4 | Q1 | Q2 | Q3 | Q4 | Q1 | Q2 | Q3 | Q4 |
| Average New Enrollments / Man Day (Thousands of People) | | 7 | 7 | 7 | 18** | 18** | 18** | 18** | 18** | 18** |
| Additional Enrollment (Millions of People) | | 0.5 | 0.5 | 0.5 | 1.2 | 1.2 | 1.2 | 1.2 | 1.2 | 1.2 |
| Cumulative Enrollment (Millions of People) | 2.3 | 2.8 | 3.3 | 3.7 | 4.9 | 6.1 | 7.3 | 8.5 | 9.7 | 10.9 |
| | | | | | | | | | | |
| Est. Total Overhead Savings (Millions of AED) | | 14 | 29 * | 43 | 80 | 116 | 153 | 190 | 227 | 264 |
| Est. Total Time Saving (Thousands of Days Productivity) | | 7.4 | 14.8 | 22.2 | 41.2 | 60.2 | 79.2 | 98.2 | 117.2 | 136.2 |

NOTE : Based on labor cost saving of 31 AED per application and wait time reduction of 23 minutes per application

\* Breakeven on cost of 200 BIWS machines acquired in 2010 to enable re-engineered process (by Q2 2011)

\*\* All visa and key government service applicants required to enroll for Emirates ID card

Fig. 7: ROI scenario 2

### 6. Lessons Learned

This section presents some of the most prominent lessons learned and consideration areas that played key roles in facilitating the overall project management.

### 6.1 Leadership and Commitment

The literature has recognized the critical role of leadership in BPR initiatives. Hammer and Champy (1993; 2003) state that most reengineering failures stem from the "breakdowns in leadership". Leadership role is seen to create a sense of mission among organizational members (Carr & Johansson, 1995; Hammer & Champy, 1993). Caron et al. (1994, p. 247) have also observed that for successful radical change, members of the senior management must be committed to the initiative, and must demonstrate their commitment "by being visibly involved with the project". See also (Gadd and Oakland, 1996; Barrett, 1994; O'Neill & Sohal, 1998).

The significant outcomes of the reengineering initiative at Emirates ID were the results of strong commitment from the Vice-Chairman, and persistent result focused top management. Business process improvement must be aligned with business objectives and clear set of outcomes. Successful implementation of change programs comes with a vision and a plan and an aggressive execution of that plan. Delegation and empowerment of teams is necessary to crease sense of responsibility of the work to be completed in such plans. This should facilitate the creation of the culture for ownership and accountability.

Cyert and March (1992), among others, point out that conflict is often a driving force in organizational behaviour. BPR claims to stress teamwork, yet paradoxically, it must be "driven" by a leader who is prepared to be ruthless. This is why top management macro and micro involvement during the execution is essential to guide and re-unite individuals and departments as conflicts would normally arise.

### 6.2 Information Technology is not a target on its own

Despite the findings of hundreds of studies which indicated it as one of the major failure reasons of initiatives, some management teams have the tendency to focus on Information Technology as a primary enabler to business needs and requirements. Top management focus to this area is crucial, as they would need to intervene at different stages to re-communicate objectives and get everybody on the same page and ensure compliance with business objectives. Hammer and Champy (1993) suggest that organizations think 'inductively' about information technology. Induction is the process of reasoning from the specific to the general which means that manager must *'learn'* the power of new technologies and *'think'* of innovative ways to alter the way work is done. This is contrary to deductive thinking where problems are first identified and solutions are then formulated. A practical approach in this area is to systematically benchmark and evaluate best practices, using relevant organizations, and consider the extent to which the processes need redesigning.

**6.3 Getting everybody on the same page**
It was difficult times to push both second and third line management teams to shift from their comfort zones and the traditional way of doing work, to think of their departments and units as components of a larger production line where performance evaluation would be based on how the entire product line is performed. There was great tendency by them to focus on their own business functions in silo environments.

The radical process reengineering introduced in the organisation required the breaking down of functional and individual job boundaries as the new processes did not have to coincide with the existing departmental structure. Internal departments were expected to be more supportive of each other and share information and best practices.

Through lots of trial and error management approaches to improve teams performance, management teams and thereafter all other employees began to realize and feel the need for change and adaptation to the new status quo. The higher management team in the organisation was needed to adopt a culture of empowerment and learning.

We map this to the story of the five blind men who attempted to define an elephant as depicted in Fig. 8. One man grabbed an ear, another the trunk, a third the tail, the fourth a leg, and the last touched the side. The man who touches the leg concludes that an elephant is thick and round and much like a column or pillar. Another man puts his hand on the trunk and concludes that an elephant is slender and flexible and must be something like a snake. The last man pushes on the elephant's side and determines that it is broad and unmovable like a large wall. Obviously what the diagram is showing here that there is no shortage of perspectives one could develop. The moral of the story is that everything is relative. Each of the blind men told the truth based on their experience with the elephant, but no one man's truth could exclude another's. No truth took precedence, even in the face of completely opposite claims.

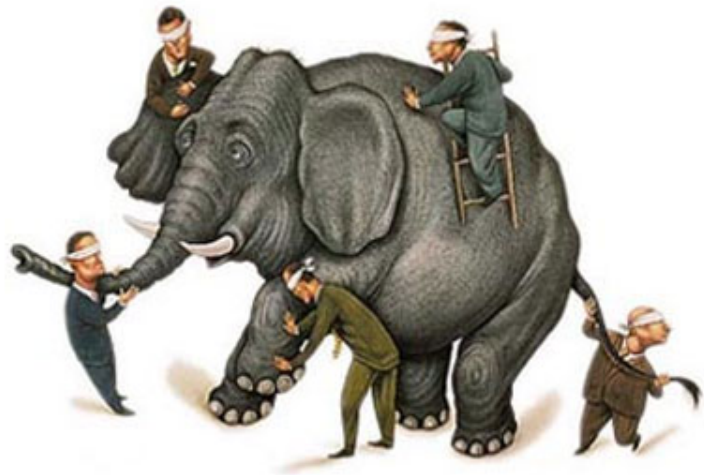

Fig. 8. Blind men and an Elephant.

This lesson is critically important for management to comprehend when introducing change in their orgnaisations. Individuals at many instances were found to work and focus on their own work within a functional area of the organization. Unless they see and comprehend that their work is part of a larger system, it would be very much challenging to get and keep everybody on the same page. For business process reengineering to succeed, lots of training and on job coaching was required to psychologically shift their mindset to the new status quo.

One improvement that helped teams to work more homogeneously was the change of work environments where all support departments were moved to a single and larger facility that facilitated improved communication between them. The second improvement was the result of the new reengineered layouts of the registration centers. The layout was radically redesigned to better suit the new registration process flow and enabled more transparency due to the glass-partition-walls of management offices and the open space layout. This also contributed to improving customer service and satisfaction.

**6.4 People and Performance Management**
Perhaps one of the most important success factors for any change program is the people element. It is important that top management make tuning decisions to create the space for change. If an organization wishes to change the way it operates, it must turn to its people to make it happen. People are the agents of change. Creating business plans and strategies are important, but they are only tools to guide the actions of people.

This should pave the way for the implementation of performance-based evaluation. The shift to performance based evaluation, management empowerment, reward for creativity, and a system-view; all helped Emirates ID to enormously improve its quality and performance for its customers. The adoption of continuous improvement philosophy led to the development of a solid proactive work environment that puts customer satisfaction and operational effectiveness and efficiency at the top of the organizational priorities.

It is important for an organisation operating in the public sector to continuously revisit its defined vision and mission to revive organisational mindset of where the organization is going, and to provide a clear picture of the desired future position. Producing key performance measures to track progress should be based on that. Management need to develop a culture for constant improvement and Identification of initiatives that will recuperate performance. Such performance management activities need to be placed in a feedback loop, complete with measurements and planning linked in Deming cycle of "Plan - Do - Check - Act", also known as the Control Circle, or PDCA. See also Fig. 9.

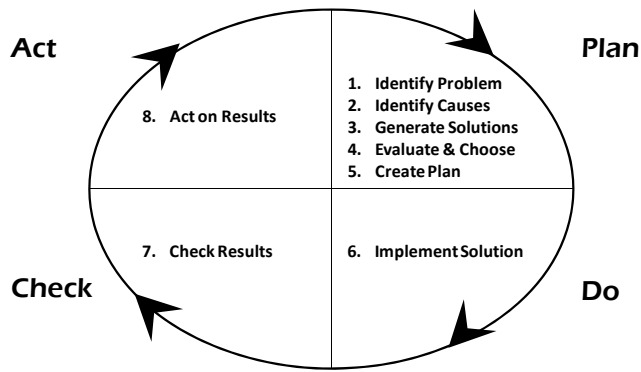

Fig 9. Deming control cycle.

**6.5 BPR and TQM: A State of Confusion in Practice**
In practice we observe organisations to have mixed or improper definitions for the application of BPR and TQM. TQM as defined in the literature is a strategic approach that is based on the premise of continuous improvement which puts emphasis on the identification of methods to continuously improve customer satisfaction, product quality, or customer service (Evans, 2004; George, 1998; Kemp, 2005). BPR on the other hand is concerned with the reorganization of the complete process cycle in major parts of the organization to eliminate unnecessary procedures, achieve synergies between previously separated processes, and become more responsive to future changes (Coulson-Thomas, 1993; Davenport, 1993). Both TQM and BPR assume that in order to provide better products and services, organizations must improve business processes.

TQM is more of a systematic approach to improving business processes through a philosophy of continuous improvement resulting in an upward sloping line of linear process improvements. BPR is not about tweaking existing processes but rather combines a strategy of promoting business innovation with radical change in business processes to achieve breakthrough improvements in products and services. See also Fig. 10 below.

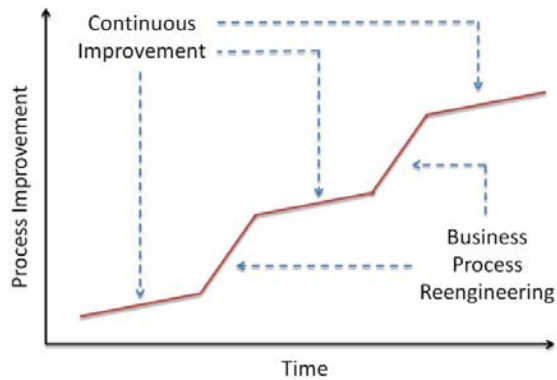

Fig. 10. TQM vs BPR. Adopted from: Hoffer et al. (2011)

It was needed that management teams to distinguish between these approaches. Unlike TQM, for instance, that aims on smoothly and incremental improvements, BPR aims on dramatic and rapid results and is suited for organisations facing gargantuan challenges to optimize the workflow and productivity. TQM targets to improve the existing systems. BPR on the other hand takes an opposite assumption as it is concerned with frame-braking change that attempts to create new systems rather than repairing old systems. BPR puts much emphasis on the enabling role of information technology and pays less attention to documentation. Table 1 provides further details about the differences between each of the two approaches.

Table 1: Comparison BPR and TQM

|  | BPR | TQM |
|---|---|---|
| **Description:** | Particular approach concerned with rethinking current systems and processes. | Concerned with improving work processes and methods in order to maximise the quality of goods and services. |
| **Type of Change:** | Planned, frame-braking | Planned, continuous |
| **Aim:** | To redefine existing work methods and processes to improve efficiency. | Keep existing customers by meeting or exceeding their expectations concerning products and services. |
| **Key Driver:** | Competitive pressures and intense need to cut costs. | Increasingly competitive market and the need to compete for specific customer demands. May also be driven by specific problems such as high costs or poor quality. |
| **Change Agent:** | External consultant | External or internal |
| **Learning process:** | Double loop | Single or double loop |
| **Nature of culture change:** | Values objectivity, control, consistency and hierarchy | Customer focused values |
| **Change to team based work:** | Yes. Requires a shift to team based work because the work is process based rather than task based. | Often requires a shift to team based work |

Source: Millett & Harvey, 1999

**6.6 Creating Sense of Agility**
Agility in public sector context is the ability of an organization to be dynamic in rapidly changing and continually fragmenting operating environments for high quality, high performance, and customer configured service models. Organisations in such environments are needed to develop information capabilities to treat masses of population as individuals and services that are perceived as solutions to their individual needs and requirements. This should help addressing the requirements of different and constantly changing customer opportunities. Goldsmith & Eggers (2004) indicate that the traditional, hierarchical model of governments simply does not meet the demands of the complex, rapidly changing era we live in, and suggests that the public sector requires agility in its systems, structures and processes. Fig. 11 depicts some pressuring elements pushing public sector organisations to adopt more agile approaches to address such requirements.

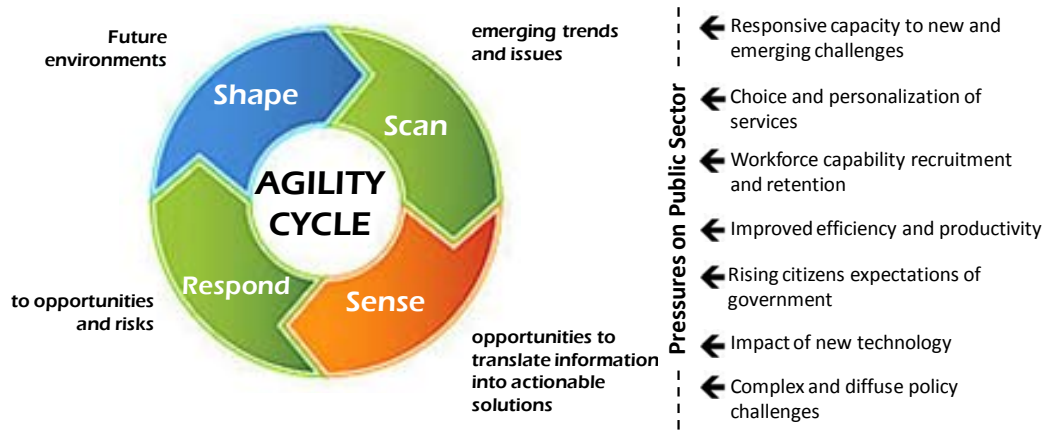

Fig. 11. Need for agility in public sector organizations.

In simple terms, we relate agility in organisations to adaptability and speed. During the BPR project, it was important to develop deep management understanding of various priority elements critical to the success of the overall project such as the organisational structure, jobs definitions, and evaluation and reward systems. This needed to be followed by an understanding of the organisation's talent and capabilities, and creating an organisational culture that supports redeployments and re-skilling. With strong and visible leadership, the organisations needed to focus on building a unified organisation that defeats silos, and developing capabilities to manage internal change well. The work presented herein attempted to add to the limited body of knowledge of practices in the field and an experience to share and build upon.

**Conclusion**

National Identity programs around the world have been going through a number of challenges. The most obvious challenge these programs face is seen to be with the quality of the enrolment process in terms of its effectiveness and efficiency of adopted processes. This study recognises the importance of this critical field, and aimed to improve overall understanding and addressing government needs for higher quality and more citizen-focused services in national ID programs.

By implementing and examining the BPR project at Emirates ID, this study provides guidelines for BPR projects in national identity initiatives with a similar context. The business process reengineering at Emirates ID resulted in substantial business benefits and contributed to the simplification of the work of the employees at front lines, increasing the degree of transparency and accuracy in functioning of the enrolment process at registration centers, and most importantly improved overall customer experience and satisfaction.

While there are similarities in how governments may approach reengineering, each government should tailor its BPR efforts to satisfy its unique conditions and operating environment (Kettinger et al., 1997). We reiterate that managing a reengineering initiative is extremely complex and difficult, and there is (and can be) no guaranteed path to success (Sauer et al., 1997; Galliers & Baets, 1998).

Although the major limitation of this research is the sample size that limits generalisability, the study is rated high on its data richness, and appropriateness for such dynamic area of practice. National Identity schemes all over the world require going through almost the same procedures with only differences related to the choice of biometric technologies adopted in each country. The lessons learned documented in this article provide practical considerations for management in the field. They are considered important building blocks for the BPR exercise to succeed.

**References**

[1] Alavi, M. and Carlson, P. "A review of MIS research and disciplinary development," Journal of Management Information Systems (8:4), 1992, pp. 45-62.
[2] Alkhouri, A.M. (2007). UAE National ID Program Case Study. International Journal Of Social Sciences, Vol. 1, No. 2, pp.62-69.
[3] Alkhouri, A.M. (2010) Facing the Challenge of Enrolment in ID Card Programs. 'The Biometric Landscape in Europe', Proceedings of the Special Interest Group on Biometrics and Electronic Signatures, BIOSIG 2010, Darmstadt, Germany, September 09 -10, 2010.

[4] Barrett, J.L., 1994. Process visualization: Getting the vision right is the key. Information Systems Management 11 (2), 14–23.
[5] Benbasat, I., Goldstein, D.K. and Mead, M. "The Case Research Strategy in Studies of Information Systems," MIS Quarterly (11:3) 1987, pp. 369-386.
[6] Caron, M., Jarvenpaa, S.L. & Stoddard, D.B. (1994, September). "Business Reengineering at CIGNA Corporation: Experiences and Lessons Learned From the First Five Years,"MIS Quarterly, pp. 233-250.
[7] Carr, D., Johansson, H., 1995. Best Practices in Reengineering. McGraw-Hill, New York.
[8] Coulson-Thomas, C.J., 1993. Corporate transformation and business process engineering. Executive Development 6 (1), 14–20.
[9] Cyert, R. M. and J. G. March (1992), A Behavioral Theory of the Firm, Oxford: Blackwell.
[10] Davenport, T.H. & Beers, M.C. (1995). "Managing Information About Processes," Journal of Management Information Systems, 12(1), pp. 57-80.
[11] Davenport, T.H. & Short, J.E. (1990 Summer). "The New Industrial Engineering: Information Technology and Business Process Redesign," Sloan Management Review, pp. 11-27.
[12] Davenport, T.H. Process Innovation: Reengineering work through Information Technology. Boston: Harvard Business School Press, 1993.
[13] Davidson, W. (1993), ``Beyond re-engineering: the three phases of business transformation'', IBM Systems Journal, Vol. 32 No. 1, Winter, pp. 65-79.
[14] Dunleavy, P., Margetts, H., Bastow, S. and Tinkler, J. (2006) New Public Management Is Dead--Long Live Digital-Era Governance. Journal of Public Administration Research and Theory 16 (3) 467-494.
[15] Earl, M., Khan, B., 1994. How new is business process redesign. European Management Journal 12 (1), 20–30.
[16] Earl, M.J., Sampler, J.L. & Short, J.E. (1995). "Strategies for Business Process Reengineering: Evidence from Field Studies," Journal of Management Information Systems, 12(1), pp. 31-56.
[17] Evans, J.R. (2004) Total Quality: Management, Organization and Strategy. South-Western College Pub.
[18] Gadd, K., Oakland, J., 1996. Chimera or culture? Business process re-engineering for total quality management. Quality Management Journal 3 (3), 20–38.
[19] Galliers, R.D., Baets, W.R.J., 1998. Information Technology and Organizational Transformation. Innovation for the 21st Century Organization. Wiley, New York.
[20] George, S. (1998) Total Quality Management: Strategies and Techniques Proven at Today's Most Successful Companies. Wiley.
[21] Goldsmith, S. & Eggers, W.D. (2004) Governing by Network: The New Shape of the Public Sector. Washington DC: Brookings Institution Press.
[22] Gordon, G.J. and Milakovich, M.E. (2009) Public Administration in America, 10 edition. Wadsworth Cengage Learning, USA.
[23] Hammer, M. and Champy, J. (1993) Reengineering the Corporation: A Manifesto for Business Revolution. New York: HarperCollins.
[24] Hammer, M. and Champy, J. (2003) Reengineering the Corporation: A Manifesto for Business Revolution. HarperCollins Publishers, NY.
[25] Hoffer J.A, George J.F, Valacich J.S (2011) Modern Systems Analysis and Design. Sixth Edition. Prentice Hall: New Jersy.
[26] Jeston, J. and Nelis, J. (2008) Business Process Management: Practical Guidelines to Successful Implementations. Second Edition. UK: Butterworth-Heinemann.
[27] Keen, P.G.W. Shaping the Future: Business Design through Information Techology. Boston: Harvard Business School Press, 1991.
[28] Kemp, S. (2005) Quality Management Demystified. McGraw-Hill Professional.
[29] Kettinger, W.J. & Grover, V. (1995). "Special Section: Toward a Theory of Business Process Change Management," Journal of Management Information Systems, 12(1), pp. 9-30.
[30] Kettinger, W.J. & Grover, V. (1995). "Special Section: Toward a Theory of Business Process Change Management," Journal of Management Information Systems, 12(1), pp. 9-30.
[31] Leedy, P. D., & Ormrod, J. E. (2005). Practical research: Planning and design (8th ed.). Upper Saddle River, NJ: Prentice Hall.
[32] Lloyd, Tom, Giant with Feet of Clay/ Tom Lloyd Offers a Contrasting View of Business Process Reengineering, Financial Times, December 5, 1994; Pg. 8.
[33] Lowenthal, J.N., 1994. Reengineering the Organization; A Step-By-Step Approach to Corporate Revitalization. ASQC Quality Press,Milwaukee, USA.
[34] Madison, D. (2005) Process Mapping, Process Improvement and Process Management. Paton Press.
[35] Millett, B. and Harvey, S. ( 1999) OD, TQM AND BPR: A COMPARATIVE APPROACH. Australian Journal of Management & Organisational Behaviour, 2(3), 30-42.
[36] O'Neill, P., Sohal, A., 1998. Business process reengineering: application and success—an Australian study. International Journal of Operations and Production Management 18 (9–10), 832–864.
[37] Orlikowski, W.J. & Baroudi, J.J. "Studying Information Technology in Organizations: Research Approaches and Assumptions", Information Systems Research (2) 1991, pp. 1-28.
[38] Porter, M.E., 1980. Competitive Strategy. Free Press, New York.
[39] Porter, M.E., 1985. Competitive Advantage. Free Press, New York.
[40] Porter, M.E., 1990. The competitive advantage of nations. Harvard Business Review 68 (2), 73–92.
[41] Sauer, C., Yetton, P.W. and associates, 1997. Steps to the Future. Fresh Thinking on the Management of IT-Based Organizational Transformation. Jossey-Bass, San Francisco, CA.
[42] Talwar, R., 1993. Business re-engineering—A strategy-driven approach. Long Range Planning 26 (6), 22–40.
[43] Yin, R. K. Case Study Research, Design and Methods, 3rd ed. Newbury Park, Sage Publications, 2002.